\documentclass[sigconf,nonacm]{acmart}

\usepackage{algorithm}
\usepackage{algpseudocode}
\usepackage{xeCJK}
\newcommand{\jp}[1]{#1}
\XeTeXlinebreaklocale "ja"
\graphicspath{{figures/}}

\title{Onomatopoeia Cursor: Verbal Mirroring of Mouse Movement with Comic-Style Lettering}

\author{Yoichi Ochiai}
\affiliation{%
  \institution{University of Tsukuba}
  \city{Tsukuba}
  \country{Japan}
}
\email{wizard@slis.tsukuba.ac.jp}

\author{Miki Okamura}
\affiliation{%
  \institution{University of Tsukuba}
  \city{Tsukuba}
  \country{Japan}
}
\email{mikio_kamura@digitalnature.slis.tsukuba.ac.jp}

\begin{document}

\begin{abstract}
The mouse cursor has remained visually mute for half a century: it shows \emph{where} we point, but says nothing about \emph{how} we move. We present the \emph{Onomatopoeia Cursor}, a shipped macOS overlay that classifies cursor kinematics in real time and renders Japanese mimetic words (onomatopoeia) as animated comic-style lettering above the cursor---\jp{キョロキョロ} (glancing around) for rapid horizontal reversals, \jp{ソロソロ} (cautiously) for slow careful motion, \jp{ビューン} (whoosh) for fast straight strokes. The system reads seven input channels---cursor kinematics, keystrokes (fact only), trackpad scrolling, clicks, drags and drops, window-edge crossings, and stillness---and displays roughly sixty word forms across five languages. Crucially, the \emph{form} of each word fluctuates with the manner of action through a morphological generator grounded in Japanese sound symbolism (voicing $=$ weight, gemination $=$ abruptness, elongation $=$ extent, reduplication $=$ iteration): a gentle click yields \jp{ポチ}, a forceful one \jp{バチンッ}; a careful drop \jp{ポトッ}, a hurled one \jp{ポーイッ！}. Beyond the rule generator, an on-device onomatopoeia-only transformer (${\approx}0.4$M parameters, dependency-free Swift/BLAS inference at ${\approx}9.7$\,ms per word) trained on 2{,}782 mimetic words conditioned on four morphology and six semantic dimensions synthesizes novel forms from the manner of an action; for events with a definite sound source (keystroke, Enter, click, edge, drop) a \emph{family-anchored} generator constrains that synthesis to the correct phonetic family. Characters are rendered with hand-drawn outline perturbation, brush-style vector deformation (pressure, taper, roughness), and per-character animation grounded in manga lettering conventions; an automated aesthetic-search loop (render $\rightarrow$ LLM judging $\rightarrow$ ridge-regressed style map) tunes the lettering. We articulate the design space of verbal motion mirroring (vocabulary, timing, typography, placement, context, modality, agency), formalize the pipeline as a learnable differentiable mapping, and report a \emph{technical} evaluation of what is actually implemented and measured---generation latency, family-constraint satisfaction, dictionary and language coverage, and an anticipatory-model feasibility sweep. Our central conjecture concerns the \emph{sense of agency}: formative first-person use by the first author suggests that naming a movement while it happens perturbs the felt authorship of the action---\emph{modulation}, \emph{amplification}, and \emph{interference}---and we lay out, as future work, a within-subjects study with salience-matched controls to test these effects. No user-study results are claimed here; the contribution is the concept, the working system, and its design space.
\end{abstract}



\maketitle

\begin{figure*}[t]
  \centering
  \includegraphics[width=0.98\textwidth]{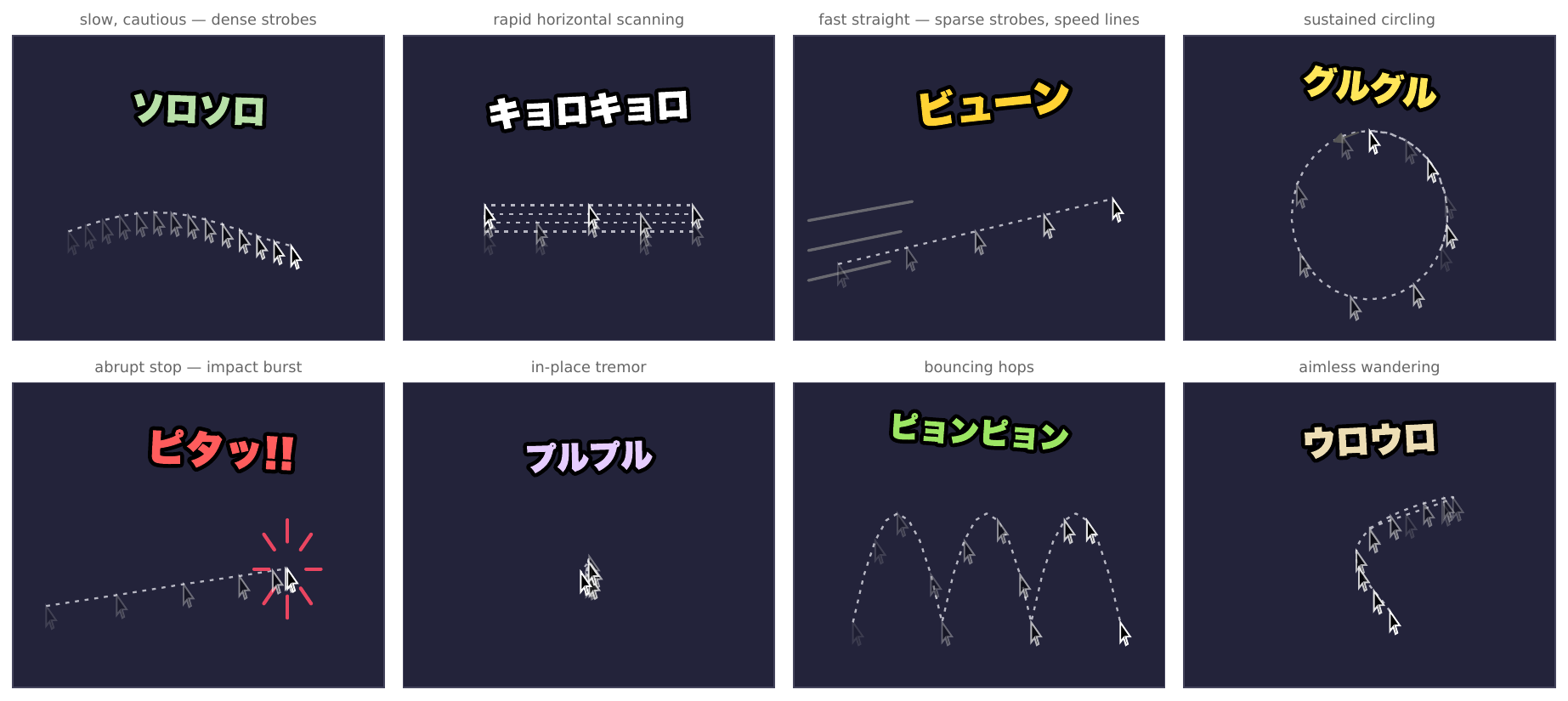}
  \caption{Eight motion words, shown as stroboscopic cursor sequences (opacity encodes time; strobe spacing encodes speed). Slow careful motion yields \jp{ソロソロ} with dense strobes; a fast straight stroke yields \jp{ビューン} with sparse strobes and speed lines; a fast stroke ending abruptly slams \jp{ピタッ!!} with an impact burst; sustained circling yields \jp{グルグル}. Word color, typeface, and animation follow manga lettering conventions.}
  \label{fig:teaser}
  \Description{Eight dark panels, each showing a dashed cursor trajectory with multiple cursor arrows fading in over time and a large manga-style Japanese onomatopoeia word: sorosoro, kyorokyoro, byuun, guruguru, pita, purupuru, pyonpyon, urouro.}
\end{figure*}

\section{Introduction}

Pointing is the most frequent act in graphical interfaces, yet the cursor conveys only position. The rich kinematic signature of how a user moves---hesitantly, playfully, frantically, smoothly---is discarded, although decades of work show that cursor dynamics carry information about intent, emotion, and cognitive state. Meanwhile, Japanese offers an unusually productive lexicon for exactly this information: mimetic words (\emph{giongo/gitaigo}) that name qualities of motion with phonetic iconicity, and a mature visual grammar---manga lettering---for placing such words next to moving bodies.

We connect these two observations. The \emph{Onomatopoeia Cursor} is a macOS overlay that continuously converts cursor kinematics into onomatopoeia rendered as animated comic-style lettering above the pointer. Slow careful movement yields \jp{ソロソロ}; rapid left-right scanning yields \jp{キョロキョロ}; a sudden stop after a fast stroke slams a red \jp{ピタッ!!} onto the screen (Figure~\ref{fig:teaser}). The words are not labels chosen by the user but a real-time \emph{verbal mirror} of their own body. The system is implemented, signed, notarized, and distributed as a working application; this paper documents its design and reports what we have been able to \emph{measure} about it, and defers the human-subjects questions to a future study.

What makes this more than a novelty is its relationship to the \emph{sense of agency} (SoA)---the feeling that ``I am the one causing this movement.'' The cursor is arguably the object over which computer users hold the strongest, most practiced SoA; it is a prosthetic fingertip. Verbal mirroring intervenes exactly there, and formative use suggests three distinguishable phenomena. \emph{Modulation}: when the system names the current movement (``you are creeping''), the felt quality of the ongoing action suddenly changes---the same physical motion acquires the named character. \emph{Amplification}: animated lettering can induce a feeling of action beyond what the finger performed---motion feels faster, bouncier, or continues subjectively when the hand has barely moved; we take this seriously enough that the shipped system reframes itself as an \emph{amplifier of the felt sensation of input} and logs the amplification directly (Section~\ref{sec:system}). \emph{Interference}: the word feeds back into motor control---and, strikingly, with inverted causality. When \jp{ピタッ!!} (an abrupt-stop word) appears, the first author reports that the cursor stopped \emph{because} the word appeared: the annotation, which in fact was triggered by the author's own deceleration, is experienced as the cause of the stop rather than its description. This postdictive reattribution---the display claims authorship of an action milliseconds after the body performed it---\emph{inverts} the priority principle of apparent mental causation (the ``cause'' arrives after the act), placing the phenomenon in the postdiction and temporal-rebinding literature rather than priority-based mental causation. We stress the epistemic status of these three phenomena: they are formative first-person observations by the first author (Section~\ref{sec:vignettes}), motivating---not established by---the study we outline as future work. The pipeline is nonetheless designed to be self-contained on screen: stimuli, conditions, and behavioral logging all live inside one distributable application, compressing the distance between prototype and controlled experiment.

\begin{figure*}[t]
  \centering
  \includegraphics[width=0.95\textwidth]{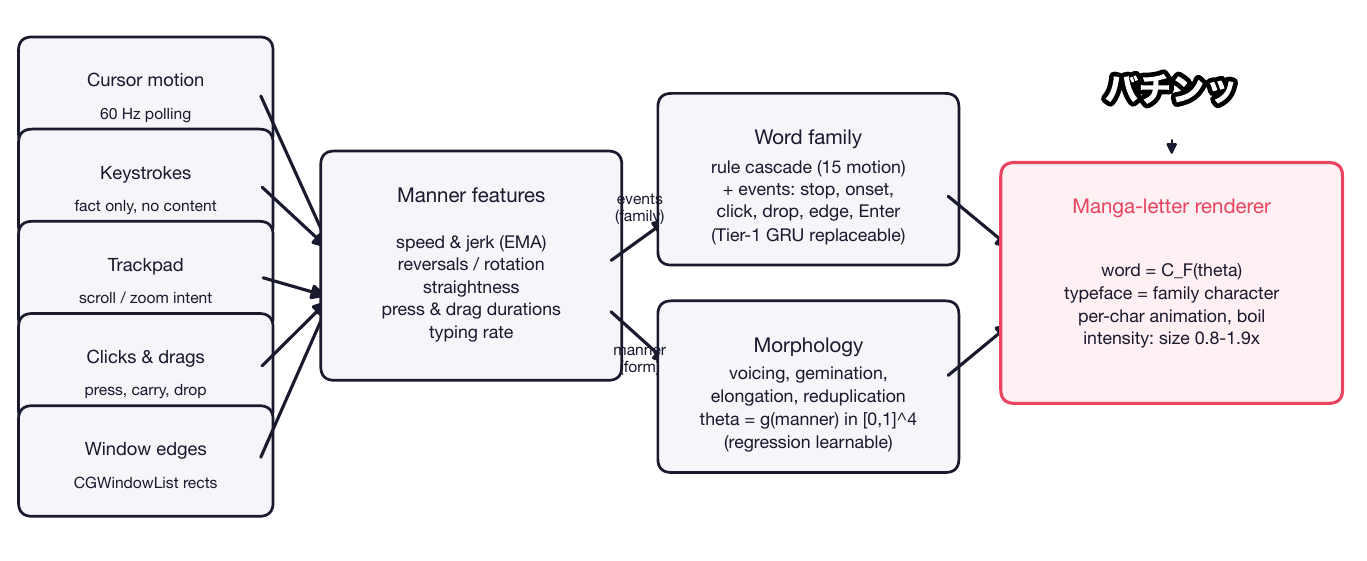}
  \caption{System pipeline. Input channels feed a shared manner-feature extractor. Events select the word \emph{family}; manner features drive the \emph{morphology} parameters $\theta$ (voicing, gemination, elongation, reduplication); the renderer composes and animates the resulting form. Both stages are replaceable by learned models (Section~\ref{sec:ml}); an on-device transformer and a family-anchored generator already stand in for parts of the closed set.}
  \label{fig:system}
  \Description{Block diagram from input channels through manner features, word family selection and morphology parameters, to a manga-letter renderer.}
\end{figure*}

This paper contributes: (1) the concept and design space of verbal motion mirroring at the cursor, framed as an intervention on the sense of agency (modulation / amplification / interference); (2) a working seven-channel system (Figure~\ref{fig:system}) whose word \emph{forms} are generated from the manner of action through a parametric model of Japanese onomatopoeic morphology (Section~\ref{sec:morph}), an on-device generative transformer, and a family-anchored generator for sound-definite events; (3) a formalization of the pipeline---manner features, family selection, morphology parameters, and hybrid low-latency/contextual inference---as a learnable, differentiable mapping; (4) a technical evaluation of the implemented system (latency, family-constraint satisfaction, coverage, and an anticipatory-model feasibility sweep) with \emph{no} simulated or user-study numbers; and (5) a discussion of verbal haptics, postdictive agency, and the OS event-visibility boundary of embodied mirroring, together with a preregisterable study design as future work.

\section{Related Work}

\subsection{Computing with onomatopoeia and sound symbolism}
Japanese mimetic words have been treated computationally as points in a continuous perceptual space: Sakamoto and colleagues quantify arbitrary onomatopoeia on dozens of affective scales from their phonology~\cite{doizaki2017}, and systematic sound--perception correspondences have been demonstrated for texture and softness~\cite{sakamoto2017tactile,nalbantoglu2024softness}. Vision systems have generated sound-symbolic words from material texture images ~\cite{yamagata2021}, sequence models map onomatopoeia to and from environmental sound ~\cite{okamoto2022onomatowave,ikawa2018}, onomatopoeic captions make non-speech sound accessible and enjoyable for deaf and hard-of-hearing viewers~\cite{kim2025onomacap}, and large vision--language models exhibit the bouba/kiki effect zero-shot ~\cite{alper2023kiki}---suggesting foundation models already carry the sound-symbolic knowledge our learned mapping (Section~\ref{sec:ml}) exploits. Closest in spirit, displaying onomatopoeia has been shown to actively modulate haptic perception ~\cite{mizoguchi2025}, supporting our hypothesis that verbal mirroring is not mere decoration; and OnomaCompass~\cite{okamura2026onomacompass} navigates between words and images for texture exploration, while sound-symbolic phoneme features have driven the \emph{generation} of dance motion from onomatopoeia~\cite{okamura2023dance}, a program unified in~\cite{okamura2026thesis}---our system extends this word$\leftrightarrow$sense program to \emph{movement} in the reading direction, inferring words from motion in real time.

\subsection{Cursor augmentation and mouse dynamics}
Cursor research has overwhelmingly optimized acquisition ~\cite{grossman2005bubble} or explained interface transitions with visual traces ~\cite{baudisch2006phosphor}. A separate literature reads cursor kinematics as a behavioral signal, estimating emotion ~\cite{yamauchi2018} and attention or intent ~\cite{arapakis2020} from trajectory features much like ours---but as silent inference for analytics, never surfaced back to the user. We close this loop: the same features that make cursor motion \emph{legible to machines} are rendered \emph{legible to the mover}.

\subsection{Comic visual language in interfaces}
Comic Chat~\cite{kurlander1996comic} demonstrated automatic translation of everyday interaction into comic form; non-photorealistic rendering has parameterized manga motion emphasis such as speed lines ~\cite{kawagishi2003cartoon}; and kinetic typography conveys affect through the motion of text itself ~\cite{forlizzi2003kinedit}. Cohn's visual-language theory~\cite{cohn2013} treats manga symbols (\emph{manpu})---including drawn onomatopoeia---as a conventionalized lexicon, and datasets such as Manga109~\cite{matsui2017manga109} make its typography learnable. Most directly, OnomatoGen~\cite{taniguchi2025onomatogen} stylizes plain text into manga onomatopoeic form (shape, size, placement) via an alpha-channel generator, and the COO dataset~\cite{baek2022coo} supplies real comic onomatopoeia for detection/recognition; these synthesize onomatopoeia \emph{visually from given text}, whereas we \emph{choose which onomatopoeia to write} from live kinematics and render it with a lightweight parametric style ($\theta\to$style)---a manga-quality generator like OnomatoGen is a natural drop-in for our renderer. Closest to our aims, Bhatia et al.\ used comic elements---onomatopoeia, speed lines, smell lines---to \emph{alter sensations} in augmented reality~\cite{bhatia2024comics}, establishing that comic visual language can change what an experience feels like, with authored overlays in head-mounted AR; and Watanabe and Yasumura's VisualHaptics generated haptic sensations from cursor visual behavior alone~\cite{watanabe2008visualhaptics}, the direct ancestor of our ``verbal haptics.'' A playful web demo by Kato renders typing and cursor onomatopoeia (\jp{カタカタ}, \jp{ポチポチ}, \jp{ビューン}) as fixed event feedback~\cite{katokatakata}, showing popular appetite for the aesthetic. We differ from all three in the same way: our words are \emph{inferred continuously from the manner of the user's own movement} (morphology included) rather than authored or event-fixed, and our question is what this inference does to the sense of agency.

\subsection{Sense of agency over cursor movements}
The sense of agency arises from the match between predicted and observed action outcomes, and is famously malleable: apparent mental causation can be induced or abolished by manipulating the timing and content of feedback ~\cite{wegner1999}, and implicit measures such as intentional binding quantify it ~\cite{haggard2002}. Cursor control is a canonical paradigm here---SoA over a moving cursor survives surprising amounts of spatial and temporal distortion, and has been studied in continuous control settings ~\cite{wen2015} and as causality perception over action--effect pairings ~\cite{kawabe2013}. Our system adds a new kind of feedback to this paradigm: not a spatial or temporal perturbation of the cursor, but a \emph{semantic annotation} of the ongoing action. Whether naming an action modulates, amplifies, or interferes with agency over it is, to our knowledge, unexplored---and uniquely testable here because the annotation is generated from the very kinematics whose ownership is in question.

\subsection{Motion-to-language models}
Mapping between movement and language is established for human bodies---joint embeddings for language-grounded pose ~\cite{ahuja2019language2pose}, motion--language corpora ~\cite{plappert2016kit}, and motion captioning; in the generative direction, onomatopoeia has served as the \emph{input} for synthesizing dance motion via the same sound-symbolic phoneme features we exploit~\cite{okamura2023dance}. Our task is a low-dimensional, real-time, stylistically constrained variant: two-dimensional cursor kinematics decoded into a mimetic lexicon and displayed within tens of milliseconds at the locus of action.

\subsection{Positioning}
Comic-language feedback exists in authored AR~\cite{bhatia2024comics} and as fixed event decorations~\cite{katokatakata}; pseudo-haptic cursors exist~\cite{watanabe2008visualhaptics}; mouse dynamics are silently inferred~\cite{yamauchi2018,arapakis2020}. To our knowledge, no prior system \emph{infers manner-graded onomatopoeic forms from the kinematics of the user's own input, system-wide and in real time}, nor frames such mirroring as an intervention on the sense of agency. We unify the four threads at the one interaction element that is always visible, shared by every GUI user, and updated at input rate.

\section{Design Space of Verbal Motion Mirroring}

We identify seven dimensions, each independently manipulable and thus usable as experimental factors.

\begin{description}
  \item[Vocabulary (what).] Closed set vs.\ open vocabulary; mimetic (\emph{gitaigo}) vs.\ onomatopoeic (\emph{giongo}); language and script.
  \item[Timing (when).] Continuous during motion; onset-triggered; event-triggered (e.g., abrupt stop); after-the-fact summary.
  \item[Typography (how).] Typeface as semantics (heavy gothic for kinetic words, mincho/serif for quiet words---a manga convention); color; per-character animation; auxiliary marks (speed lines, focus lines, outline boil).
  \item[Placement (where).] At the cursor; along the trail (afterglow); at the text caret; at screen periphery.
  \item[Context sensitivity.] Kinematics only vs.\ content-aware (what is under the cursor) vs.\ application-state-aware.
  \item[Modality.] Visual lettering; sound effects; haptics.
  \item[Agency.] Mirroring one's own cursor vs.\ verbalizing another's (screen sharing, collaborative editing).
\end{description}

\begin{figure*}[t]
  \centering
  \includegraphics[width=0.98\textwidth]{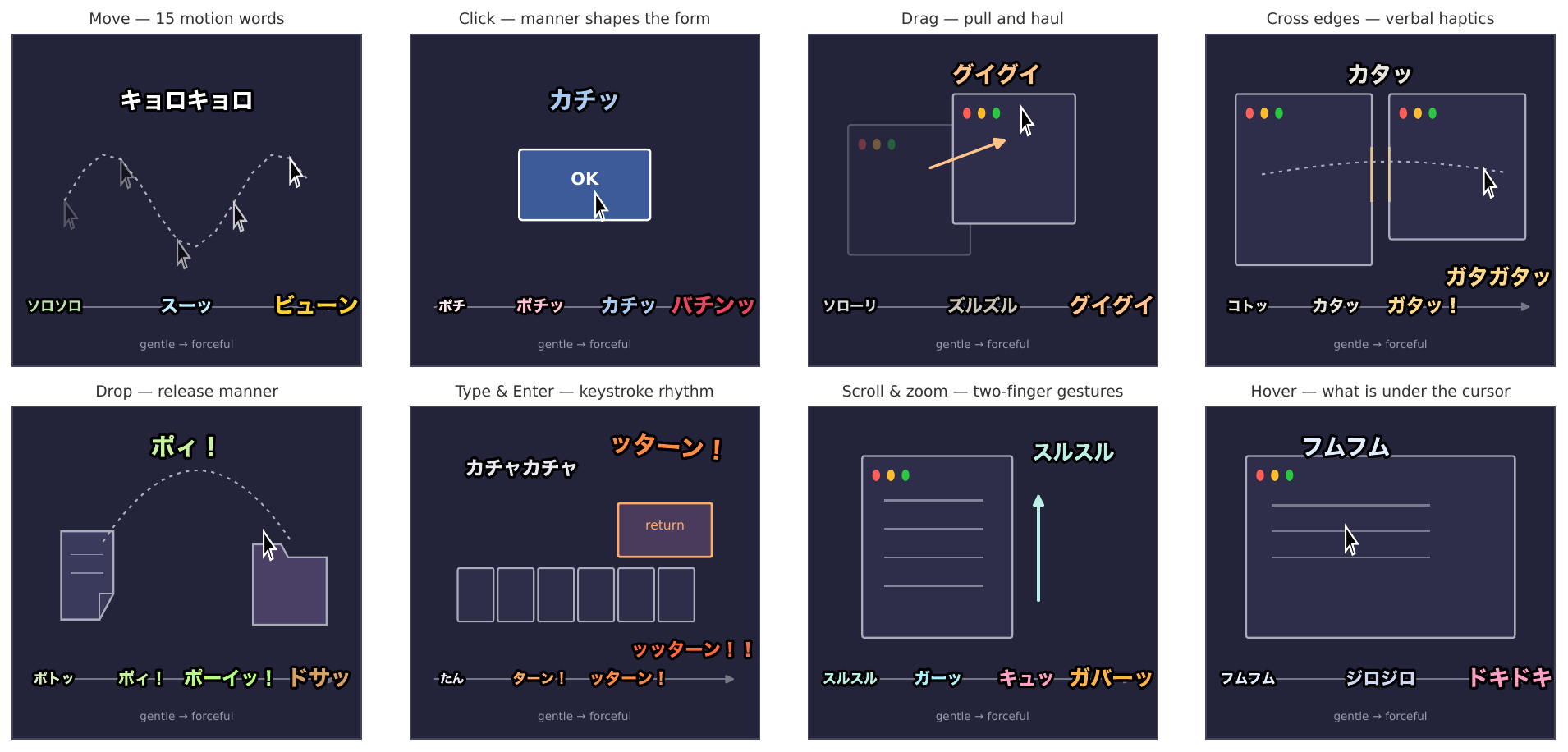}
  \caption{The interaction design space. Each panel is one everyday interaction with its word family; the strip below each panel shows how the \emph{form} fluctuates with the manner of execution (gentle $\rightarrow$ forceful), following the sound-symbolic morphology of Section~\ref{sec:morph}.}
  \label{fig:designspace}
  \Description{Eight illustrated panels on dark backgrounds: moving the cursor, clicking an OK button, dragging a window, crossing window edges, dropping a file into a folder, typing and pressing return, scrolling a page, and hovering over text. Each panel shows a manga-style Japanese word and a gradient strip of word forms from gentle to forceful.}
\end{figure*}

Our current system instantiates: roughly sixty word forms over seven input channels in five languages, morphology-parameterized vocabulary (Section~\ref{sec:morph}), continuous + event timing, manga-grounded typography, cursor and caret placement, optional element-level context, visual modality, self-mirroring. Figure~\ref{fig:designspace} lays out the interaction scenarios: each everyday act---moving, clicking, dragging a window, crossing its edge, dropping a file, typing, scrolling, hovering, and holding still---carries a family of forms spanning gentle to forceful execution.

\section{The Onomatopoeia Cursor System}
\label{sec:system}

The following describes the system as currently shipped. It is a menu-bar macOS application (Apple Silicon and Intel, macOS 12+), click-through, following the cursor across spaces and full-screen apps.

\subsection{Kinematic features and classifier}
The overlay polls the global cursor position at 60\,Hz and computes exponentially smoothed speed and jerk, horizontal/vertical direction-reversal counts (0.6\,s window), cumulative rotation of the velocity vector (0.7\,s), and straightness (net displacement over path length, 0.45\,s). A prioritized rule cascade maps these to fourteen motion words, and an edge detector fires \jp{ピタッ!!}---the fifteenth motion word---when a fast stroke ends abruptly (Table~\ref{tab:words}) (with a refractory period to avoid over-triggering, a threshold tuned through formative use). The rule classifier serves three roles: a baseline condition, a stable demo, and a weak-supervision teacher for learned models.

\begin{table}[h]
  \caption{Motion vocabulary and kinematic conditions.}
  \label{tab:words}
  \footnotesize
  \setlength{\tabcolsep}{2.5pt}
  \begin{tabular}{lll}
    \toprule
    Word & Gloss & Kinematic signature \\
    \midrule
    \jp{グルグル}/\jp{クルクル} & spinning & high/med.\ cumulative rotation \\
    \jp{ブンブン}/\jp{キョロキョロ} & swing/glance & horiz.\ reversals; high/med.\ speed \\
    \jp{ピョンピョン} & hopping & vertical reversals \\
    \jp{プルプル} & trembling & many reversals, low amplitude \\
    \jp{ダダダッ}/\jp{ビューン} & dashing/whoosh & high speed, rough vs.\ straight \\
    \jp{スーッ}/\jp{スイスイ} & gliding & smooth straight vs.\ smooth curved \\
    \jp{ウロウロ} & wandering & mid speed, low straightness \\
    \jp{オロオロ} & flustered & high jerk \\
    \jp{ソロソロ}/\jp{ジリジリ} & cautious/creeping & low/very low speed \\
    \jp{ピタッ!!} & abrupt stop & fast$\rightarrow$still edge (event) \\
    \bottomrule
  \end{tabular}
\end{table}

\subsection{Context sensing}
An optional mode queries the accessibility API (5\,Hz) for the role of the element under the cursor. At low speeds, context words override quiet motion words: hovering text yields \jp{フムフム} (reading, nodding), images \jp{ジロジロ} (scrutinizing), buttons and links \jp{ドキドキ} (heartbeat). This is deliberately minimal---a probe of the context axis---while the full content-aware mapping is delegated to the learned model (Section~\ref{sec:ml}).

\subsection{Input channels beyond the pointer}
Interaction is verbal wherever the body touches the machine. Keystrokes produce typing words and Enter words (using only the \emph{fact} of a keypress and whether it is Return---never which key); trackpad scrolling is voiced by direction and momentum (\jp{スルスル}, \jp{ガーッ}, \jp{シュルシュル}, \jp{シャッ}); clicks, drags, and drops form a narrative (\jp{ポチッ} $\rightarrow$ \jp{グイグイ} $\rightarrow$ \jp{ポィ！}); crossing a window boundary produces \jp{カタッ}---a \emph{verbal haptic} that gives the flat screen a felt topography, computed from window rectangles alone; and \emph{stillness} is voiced too---after five seconds at rest the cursor breathes \jp{すぅすぅ}, \jp{ウトウト}, \jp{ぼー…}, occasionally startling awake with \jp{ビクビクッ}. Typing words are placed at the text caret (via the accessibility API) rather than the abandoned mouse position, with a validated fallback to the focused window when a caret rectangle is unavailable (e.g., Electron apps). One channel is architecturally closed: macOS synthesizes pinch and rotation gestures only inside the focused application, so no public API can observe them system-wide; we substitute the modifier-key zoom convention and treat this as a finding about how the OS delimits which bodily data can be mirrored at all (Section~\ref{sec:discussion}).

\subsection{Morphological generation: manner determines form}
\label{sec:morph}
An event determines only the word \emph{family}; the manner of the action determines its \emph{form}. Japanese sound symbolism provides a ready-made parametric space (Figure~\ref{fig:morph}a): voicing (\jp{゛}) encodes weight and force (\jp{カタッ}$\rightarrow$\jp{ガタッ}), gemination (\jp{ッ}) encodes abruptness (\jp{ターン}$\rightarrow$\jp{ッターン！}), elongation (\jp{ー}) encodes spatial or temporal extent (\jp{ポイ}$\rightarrow$\jp{ポーイ}), and reduplication encodes iteration (\jp{カチ}$\rightarrow$\jp{カチカチ}). Formally, each event $e$ of family $F$ carries a manner vector $u_e$ (approach speed, press duration, carry duration, release speed, typing rate, crossing speed, \ldots), mapped to morphology parameters
\begin{equation}
\theta = (\theta_{\mathrm{voi}}, \theta_{\mathrm{gem}}, \theta_{\mathrm{len}}, \theta_{\mathrm{rep}}) = g(u_e) \in [0,1]^4,
\qquad w = \mathcal{C}_F(\theta),
\label{eq:morph}
\end{equation}
where $\mathcal{C}_F$ composes the family stem with the (quantized) parameters into a surface form. The shipped $\mathcal{C}_F$ (\texttt{MorphSynth}) selects a stem along a voicing ladder, then rule-composes gemination (family-specific prefix/suffix), elongation, reduplication, and exclamation marks; because $\theta$ is continuous it can synthesize forms \emph{between} the existing rungs of a gradient (e.g.\ a form halfway between \jp{ポチッ} and \jp{カチッ}). The actual behavior of the rule generator is shown in Figure~\ref{fig:morph}b--c. Because $\theta$ is continuous, $g$ is directly replaceable by a regressor $g_\phi$ trained first by distilling the rules and then from human preference sliders over $\theta$---making word-form generation differentiable and smoothly extensible to open vocabulary. This reframing---from ``event triggers word'' to ``manner shapes form''---is what makes the inference feel \emph{apt}, and aptness is the hypothesized active ingredient of the agency effects: a word that fails to match the manner of an action has no purchase on its ownership.

\subsection{On-device generative onomatopoeia}
\label{sec:onomaformer}
Beyond interpolating between shipped forms, an \emph{onomatopoeia-only} character-level transformer, \textsc{OnomaFormer} (${\approx}0.4$M parameters), generates genuinely novel forms outside the training lexicon at a requested condition. It is trained on 2{,}782 Japanese mimetic words---assembled by growing the corpus $384\rightarrow604\rightarrow2{,}782$, the largest tier drawn from the JMDict \emph{on-mim} tag (2{,}639 representations, CC~BY-SA~4.0)---conditioned on the four morphology dimensions \emph{and} six semantic dimensions (impact, motion, texture, emotion, light, wet/dry), for a ten-dimensional condition vector; 906 dictionary entries were auto-tagged with semantics from their glosses. In \emph{generation mode} the classifier's word is replaced by an OnomaFormer form synthesized from the current motion's $\theta$ plus the channel/element's semantic category (typing $\to$ impact, dragging $\to$ motion, text hover $\to$ emotion, \ldots), so the same movement yields a slightly different fresh word each time---the ``variation of the input act.'' The model is embedded on-device as a dependency-free pure-Swift forward pass over Accelerate/BLAS (a \texttt{norm\_first} transformer with causal masking and temperature sampling), requiring no \texttt{coremltools}; it adds ${\approx}9.6$\,MB to the app and runs at ${\approx}9.7$\,ms per word with a $\theta$-bucket cache. We verified it against the PyTorch reference to near-parity (a float-accumulation difference in a single word tolerated).

\subsection{Family-anchored generation for sound-definite events}
\label{sec:family}
Unconstrained generation is right for cursor \emph{motion}, which has no physical sound source; but events with a definite source---keystroke, Enter, click, window edge, drop---have a phonetic \emph{family} fixed by their sound symbolism (a keystroke is a \jp{カタカタ}-type percussion; Enter a single \jp{ターン}/\jp{かーん}-type strike). Free generation there occasionally emitted words unrelated to the act. The family-anchored generator (\texttt{FamilyGen}) keeps generative freedom \emph{inside the family}: (1)~$\theta_{\mathrm{voi}}$ selects a stem seed from the family's voicing ladder (light$\to$heavy); (2)~OnomaFormer is forced to begin with that seed and samples only the continuation (\jp{カタ}$\rightarrow$\jp{カタカタッ！}); (3)~if the result leaves the family's admissible character set it is retried, and finally falls back to \texttt{MorphSynth}, which composes from the family table and therefore always returns a family-faithful form. This encodes the first author's sound-symbolic intuition as a hard constraint: the family is fixed for sound-definite events, and only cursor motion generates freely.

\subsection{Manga-grounded rendering}
Words are drawn per character. Each glyph is converted to its vector outline, perturbed by deterministic noise into three variants cycled at 8\,fps---the classic animation ``boil'' that makes lettering read as hand-drawn---and further deformed by a brush model (\texttt{BrushDeform}) that swells and thins the outline along its normal (pressure), pinches the entry and exit of each stroke (taper), and carves faint scratch lines through the glyph (roughness). Each word has a typeface (heavy display gothic for kinetic words, display mincho for quiet ones), a color, and a word-specific per-character animation (staggered pop-in with overshoot; swing, jitter, spin, shear with speed lines, squash-and-stretch hops, heartbeat pulses), plus BOOM-style comic effects (starbursts and halftone for impact words, diagonal flash for speed words, cloud balloons for quiet ones). Character size scales with movement intensity ($0.8$--$1.9\times$). The style parameters (typeface weight, size, tilt, hue, shadow, speed lines, and the brush pressure/taper/roughness) are themselves a function of $\theta$: heavy words are drawn heavy (\jp{ガタッ} large, red, deep-shadowed, forward-leaning), light ones light (\jp{すぅー} small, pale blue). The overlay window is click-through; words fade after 1.3\,s of stillness. Following second-author feedback---words can switch too fast to read during rapid motion, and the comic styling can overload busy backgrounds---the shipped system also provides a \emph{high-legibility} display variant (white fill with a dark gray drop shadow, comic effects suppressed) and scales the minimum word-hold time with movement intensity; both double as display-style conditions for the future study (Section~\ref{sec:future}).

\subsection{Aesthetic-search loop for the lettering style}
\label{sec:aesthetic}
The $\theta\to$style map $h(\theta)$ is refined by an automated self-improving loop, packaged as an unattended nightly script (scheduling is operator-controlled). A headless renderer (\texttt{onomatope-render}, no screen or permissions, CI-capable) produces batches of candidate letterings; a vision LLM scores each on seven axes (manga-ness, readability, meaning fit, impact, weight fit, color, overall), discriminating between candidates of the same word; and \texttt{fit\_style\_map.py} ridge-regresses the residual of the winning candidates against the hand-tuned rule map, writing $h(\theta)=h_0(\theta)+\Delta(\theta)$ to a style-map file that the app loads at launch (\texttt{StyleMap}). With sparse data the regularizer keeps $\Delta\approx0$ (the rule map, safe); as scores accumulate, the judge's taste seeps into the mapping over seventeen clamped style fields. This is a self-driving-laboratory pattern applied to the \emph{look} of the lettering, complementing the self-driving morphology study (Section~\ref{sec:formal}) that does the same for word \emph{forms}.

\subsection{Amplification logging}
\label{sec:amplog}
The system reframes itself as an \emph{amplifier of the felt sensation of the input act}: the \jp{カタカタ} of typing amplifies the sense of typing, the \jp{ビューン} of a cursor amplifies the sense of speed. To measure this without asking the user anything, an amplification log records, at each word onset, the act intensity just before (\emph{pre}) and the peak intensity in the following $0.8$\,s (\emph{post}), storing $\mathrm{amp}=\mathrm{post}/\mathrm{pre}-1$ keyed by (word, $\theta$, channel). This turns ordinary use into an implicit-reward signal---which words amplify which acts, and by how much---for steering the generator toward amplification rather than mere congruence.

\subsection{Research logging and grounding data}
An opt-in logger writes per-frame kinematic features, the displayed word, and (in context mode) the element role to plain JSONL---no screen content, no keystroke content. A menu-bar indicator makes recording state always visible. In-app mechanisms also collect supervised signal: a ``that's wrong'' gesture (double right-click) records negatives, a rare two-alternative ``which is more natural?'' probe records Bradley--Terry preference pairs, and a ``summon'' mini-game records the strongest positive examples (a movement a user made \emph{in order to} produce a named word). The same log formats feed model training, annotation, and experiment analysis.

\section{Formal Model, Latent Spaces, and Objectives}
\label{sec:formal}

\subsection{Manner features}
Let $p_t \in \mathbb{R}^2$ be the cursor position sampled at $f{=}60$\,Hz and $\Delta p_t = p_t - p_{t-1}$. Smoothed speed and jerk are exponential moving averages with $\alpha{=}0.15$:
\begin{align}
\bar v_t &= (1{-}\alpha)\,\bar v_{t-1} + \alpha \lVert \Delta p_t \rVert,
\label{eq:ema}\\
j_t &= (1{-}\alpha)\, j_{t-1} + \alpha \,\bigl|\, \lVert\Delta p_t\rVert - \lVert\Delta p_{t-1}\rVert \,\bigr|. \nonumber
\end{align}
Cumulative rotation and straightness over sliding windows $W_\tau$ (of $\tau$ seconds) are
\begin{align}
\Theta_t &= \!\!\sum_{i \in W_{0.7}}\!\! \angle(\Delta p_{i-1}, \Delta p_i)\;
\mathbb{1}\!\left[\langle \Delta p_{i-1}, \Delta p_i\rangle > 0\right],
\label{eq:turn}\\
S_t &= \lVert p_t - p_{t - 0.45 f} \rVert \Big/ \textstyle\sum_{i \in W_{0.45}} \lVert \Delta p_i \rVert, \nonumber
\end{align}
where the indicator excludes near-reversals (else reciprocating motion random-walks into spurious rotation). With reversal counts $n^x_t, n^y_t$ (0.6\,s window) and path length $L_t$, the manner vector is $m_t = (\bar v_t, j_t, n^x_t, n^y_t, \Theta_t, S_t, L_t)$; interaction channels append press duration $\tau_p$, carry duration $\tau_d$, release speed $v_r$, typing rate $r_k$, and crossing speed.

\subsection{Family selection, morphology, and anchoring}
The motion word is a prioritized cascade $w_t = \rho(m_t) = r_{k^*}$, $k^* = \min\{k : m_t \in R_k\}$ over fourteen regions; discrete events $e \in \{\textrm{stop, onset, click, drop, edge, Enter}\}$ fire as predicates on raw kinematic edges (e.g., stop $=\mathbb{1}[\bar v$ recently ${>}22 \wedge \lVert\Delta p\rVert{<}0.6$, in px/frame at 60\,Hz, for two frames$]$, with a refractory period). An event selects a family $F(e)$ with stem $s_F$; its manner subvector $u_e$ is mapped to morphology parameters $\theta$ and composed into a surface form by Eq.~\ref{eq:morph}. For sound-definite families the composition is \emph{anchored}: $\theta_{\mathrm{voi}}$ selects a seed on the family voicing ladder, the generative model is constrained to that seed and to the family's admissible character set $\Sigma_F$, and any escape falls back to the deterministic $\mathcal{C}_F$ (Section~\ref{sec:family}). The rule $g$ is piecewise-constant (Figure~\ref{fig:morph}b--c); its learned replacement is fit in two stages,
\begin{align}
\mathcal{L}_{\mathrm{dist}}(\phi) &= \mathbb{E}_{u}\,\bigl\lVert g_\phi(u) - g_{\mathrm{rule}}(u) \bigr\rVert^2,
\label{eq:morphloss}\\
\mathcal{L}_{\mathrm{pref}}(\phi) &= -\!\!\sum_{(a \succ b)}\!\! \log \sigma\!\left(s_\phi(\theta_a \mid u) - s_\phi(\theta_b \mid u)\right), \nonumber
\end{align}
distillation from the rule teacher followed by preference learning (Bradley--Terry over pairs of forms shown for the same action, or direct slider annotation of $\theta^\ast$). Because $\theta$ is continuous, word-form generation is differentiable end-to-end.

\subsection{A self-driving morphology study.}
Because $\mathcal{C}_F(\theta)$ is generative, the system can synthesize word forms it has never shipped---forms \emph{between} the existing rungs of a gradient---and enter them into in-situ two-alternative preference probes against established forms, blind to the participant. Preferences are fit by Bradley--Terry; a synthesized form that beats its established neighbors (win rate $>0.5$, $n \ge 5$) is proposed for human approval and then delivered over the air to all installations, where it immediately faces new comparisons. The apparatus thus generates its own hypotheses (candidate forms), runs its own blinded experiments, and revises its own grammar---a self-driving-laboratory pattern applied to an interaction vocabulary. The mechanisms (in-situ probe, Bradley--Terry fitting, human approval, over-the-air delivery) are implemented, and probes run only under the research-participation consent gate of Section~\ref{sec:ethics}; at the time of writing the installed base is the development team, so fleet-scale comparisons have not yet accumulated. Candidates from OnomaFormer that round-trip through a phonological analyzer to the requested $\theta$ enter the same blinded arena, making the transformer the generative source upstream of the rule interpolator.

\subsection{Anticipatory latent code}
Tier~1 replaces $\rho$ with a recurrent encoder over raw increments $x_k = (\Delta x_k, \Delta y_k, \bar v_k, j_k)$:
\begin{align}
h_k &= \mathrm{GRU}_\phi(x_k, h_{k-1}) \in \mathbb{R}^{64},
\qquad
z_K = \tfrac{1}{K} \textstyle\sum_{k \le K} h_k,
\label{eq:gru}\\
\hat y &= \mathrm{softmax}(W z_K + b), \nonumber
\end{align}
trained with class-weighted cross-entropy against the settled rule label of the \emph{full} window,
$\mathcal{L}_{A}(\phi) = -\,\mathbb{E}\left[\, \omega_{y} \log \hat y_{y} \,\right]$ with $\omega_c \propto 1/\mathrm{freq}(c)$---an early-classification distillation whose accuracy--latency frontier, measured by actually training at $K \in \{6,12,24,45,90\}$, is Figure~\ref{fig:ml}. Implementation: 2-layer GRU, hidden 64, window 90 frames, trained on 30 synthetic sessions (108k frames) labeled by the real classifier, Adam $10^{-3}$; a Core~ML conversion path is prepared for on-device deployment (not yet shipped).

\subsection{A shared motion--word latent space}
A latent-space mode embeds words by their phonology and morphology, $e_w = \psi(\mathrm{phon}(w), \theta(w))$, and aligns them with motion codes $z$ by a contrastive objective,
\begin{equation}
\mathcal{L}_{\mathrm{NCE}} = -\,\mathbb{E}_i \left[ \log
\frac{\exp(\langle z_i, e_{w_i}\rangle / \tau)}{\sum_j \exp(\langle z_i, e_{w_j}\rangle / \tau)} \right],
\label{eq:nce}
\end{equation}
yielding (i) open-vocabulary retrieval (nearest word to any motion), (ii) \emph{gap detection}---motion regions far from every existing word are sites for new-word generation, surfaced live in the latent-space HUD---and (iii) annotation-by-navigation, where a user steers through the space to find the apt form.

\subsection{The comfort objective}
Arbitration is not accuracy maximization but \emph{felt-quality} maximization. A display policy $\pi$ chooses $W_t$ among tier candidates $\{(w_i, c_i, \ell_i)\}$ (word, confidence, latency); we score a displayed sequence by
\begin{align}
J(\pi) = \mathbb{E}\bigl[\, \kappa_a \,\mathrm{Apt} - \kappa_s\, \mathrm{Switch}
- \kappa_d (\mathrm{Dens} - d^\ast)^2
+ \kappa_n\, \mathrm{Nov} \,\bigr],
\label{eq:comfort}
\end{align}
aptness (congruence ratings), stability (flicker), rhythm (word density near a target $d^\ast$, the ``panel pacing'' of manga), and a bounded novelty bonus for rare forms. The shipped heuristic---fastest-first, confidence-margin override, hysteresis, event interrupts---is a hand-tuned point in this policy space; $\pi_\psi$ is to be learned from congruence ratings plus one-tap annoyance reports as negative signal.

\begin{algorithm}[t]
\caption{Per-frame pipeline (60\,Hz)}
\label{alg:tick}
\begin{algorithmic}[1]
\State $f \gets \textsc{MotionTracker.update}(p_t)$ \Comment{Eqs.~\ref{eq:ema}--\ref{eq:turn}; teleport guard}
\State $I_t \gets \mathrm{EMA}\bigl(\max(\bar v_t/32,\; r_k/9,\; \mathrm{scroll}/40)\bigr)$ \Comment{intensity $\to$ size $0.8{+}1.1 I_t$}
\ForAll{pending events $e$ (stop, onset, click, drop, edge, Enter)}
    \State $u_e \gets \textsc{Manner}(e, f)$;\; $\theta \gets g(u_e)$
    \State $w \gets \textsc{FamilyGen}(F(e), \theta)$ \textbf{or} $\mathcal{C}_{F(e)}(\theta)$ \Comment{Eq.~\ref{eq:morph}, Sec.~\ref{sec:family}}
    \State \textsc{Display}$(w,\ \mathrm{interrupt})$;\; log event;\; amp-log $(w,\theta,\mathrm{ch})$
\EndFor
\If{$t \ge \mathrm{holdUntil}$} \Comment{hysteresis}
    \State $w \gets \rho(m_t)$ or \textsc{OnomaFormer}$(\theta,\mathrm{cats})$ (generation mode)
    \State \textsc{Display}$(w)$ \textbf{if} margin \& semantic distance warrant \Comment{Eq.~\ref{eq:comfort} heuristic}
\EndIf
\State render per-char: pop-in $\circ$ word animation $\circ$ boil $\circ$ brush$(\theta)$, scaled by $I_t$
\State \textbf{if} logging \textbf{then} append $(m_t, w, \mathrm{events})$ to JSONL
\end{algorithmic}
\end{algorithm}

\section{Technical Evaluation}
\label{sec:eval}

\begin{figure*}[t]
  \centering
  \includegraphics[width=0.98\textwidth]{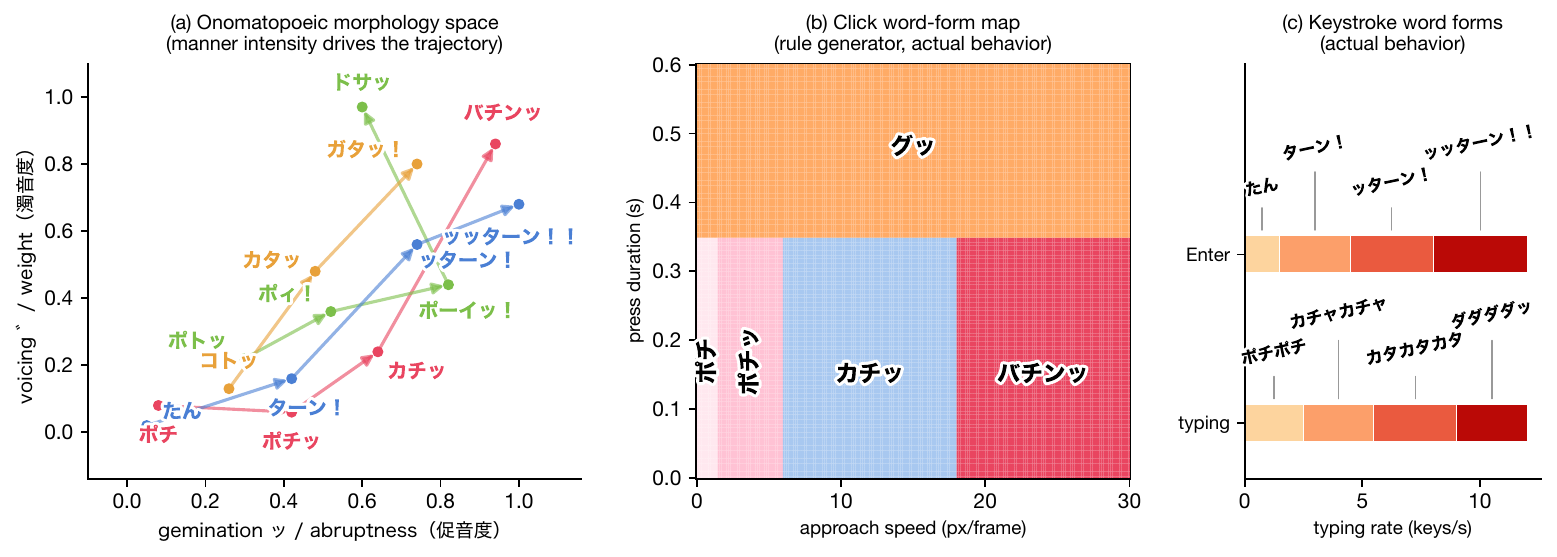}
  \caption{Onomatopoeic morphology (actual system behavior). (a)~Word families traverse the sound-symbolic parameter plane (gemination $\times$ voicing) as manner intensifies. (b)~Click word-form map over approach speed and press duration, and (c)~keystroke word forms over typing rate---both show the \emph{implemented} rule generator, which serves as the weak teacher for the learned regressor $g_\phi$.}
  \label{fig:morph}
  \Description{Three panels: a scatter of Japanese onomatopoeia arranged in a two-dimensional sound-symbolic space with arrows showing family trajectories; a two-dimensional decision map of click word forms; and horizontal bars showing typing and Enter word forms by typing rate.}
\end{figure*}

We report only what the implemented system lets us \emph{measure}; no human-subjects data are claimed. All figures in this section reflect actual system behavior (Figure~\ref{fig:morph}) or real training runs (Figure~\ref{fig:ml}).

\textbf{Generation latency.} OnomaFormer runs entirely on-device as a dependency-free Swift/BLAS forward pass (no \texttt{coremltools}), adding 9.3\,MB to the app, and generates a novel word in ${\approx}9.7$\,ms with a $\theta$-bucket cache---well within the 60\,Hz frame budget, so generation mode never stalls the overlay. Its output matches the PyTorch reference to near-parity (a single word differing by float-accumulation).

\textbf{Family-constraint satisfaction.} For sound-definite events the family-anchored generator must never leave its phonetic family. Regression tests exercise the seeded generation path repeatedly---including a high-temperature, long-seed stress test of hundreds of consecutive generations---and every accepted word lies within the family's admissible character set, with the deterministic \texttt{MorphSynth} fallback guaranteeing a family-faithful form when sampling escapes. A prior out-of-range crash in seeded position embeddings (introduced with family anchoring) was fixed and locked down by this stress test.

\textbf{Morphology behavior.} Figure~\ref{fig:morph}b--c shows the \emph{actual} decision surface of the rule generator over click approach-speed$\times$press-duration and over typing rate---the weak teacher for the learned regressor $g_\phi$. The generator is monotone in each $\theta$ dimension by construction, verified by unit tests.

\textbf{Coverage.} The system displays roughly sixty word forms over seven input channels in five languages (Japanese, English, Chinese, Korean, French), each redesigned in the target language's sound-symbolic devices rather than translated, with every form annotated by $\theta=[\text{voicing},\text{gemination},\text{elongation},\text{reduplication}]$ (the language packs double as multilingual ML supervision). OnomaFormer is trained on 2{,}782 mimetic words (the largest tier from the JMDict \emph{on-mim} tag, 2{,}639 representations).

\textbf{Anticipatory-model feasibility.} As an honest feasibility check---not a claimed win---we trained the anticipatory GRU (Eq.~\ref{eq:gru}) by truncating training windows to their first $K$ frames on synthetic trajectories labeled by the rule classifier (30 sessions, 108k frames). On this distillation data, early-classification accuracy over the thirteen reachable classes (chance $7.7\%$; of the fifteen motion words, \jp{ピタッ!!} is event-only and one rare word never wins a 90-frame majority label in the synthetic distribution) rises from ${\approx}46\%$ at a 100\,ms input window to ${\approx}73\%$ at 1.5\,s (Figure~\ref{fig:ml}). Two caveats follow. First, the x-axis is input-window length, not end-to-end system latency. Second, at the Tier-1 operating band (${\sim}200$\,ms) accuracy is below $0.5$---so whether a learned model can beat the rule cascade's structural latency floors \emph{on real human data} is an open empirical question, not a demonstrated result.

\textbf{Engineering.} The core (feature tracker, classifier, morphology generator, family-anchored generator, OnomaFormer inference, research log, vocabulary store) is AppKit-independent and covered by an XCTest-independent test runner (74 tests as of the current release). Static analysis (Snyk Code) reports no high- or medium-severity findings; two low-severity notes on a developer-only synthetic-data CLI were reviewed and accepted. The app is Developer-ID signed and Apple-notarized so downloads open without a first-run bypass.

\section{Formative First-Person Accounts}
\label{sec:vignettes}

Following first-person research practice, we report formative accounts from both authors: dated vignettes from the first author's use on 2026-07-16 (the first day the packaged app ran with the research log on; earlier unlogged development use also shaped the design), and the second author's notes after extended use. These are motivating observations that shaped the design, \emph{not} evidence; they are what the future study (Section~\ref{sec:future}) is designed to test.

\textbf{Modulation (first launch of the packaged app, 09:21).} ``I can't open it\ldots got it now (laughs). \emph{It feels strange.} This is unmistakably a new kind of interaction''---the strangeness preceding any vocabulary knowledge, arising from the mirroring itself.

\textbf{Amplification / summoning (10:30).} ``Once I knew \jp{グルグル} existed I found my hand drawing circles to summon it---the vocabulary pulls movement out of the body. When the word is bigger than the movement deserved, the movement feels bigger too.''

\textbf{Interference / inverted causality (16:00).} ``It is not that being watched makes me want to stop---when \jp{ピタッ!!} appears, it feels like the cursor stopped \emph{because} the word said so. My own deceleration triggered the word, but the word reads as the cause.'' The log shows the stop preceding word onset by ${\approx}40$\,ms on each such event in that day's log.

\textbf{Second-author observations (2026-07-29).} After extended use, the second author's notes independently separate the two constructs of the future study: on the \emph{ownership} side, ``one's own movement acquires an outline and one becomes self-aware of it---the feeling of being coached could lead to behavior change''; on the \emph{agency} side, ``the pointer itself acquires a character, and one feels one is moving \emph{through} it---close to operating a game avatar.'' She also noted an asymmetry in the vocabulary: words like \jp{カチャカチャ} or \jp{ピタッ} could equally describe the hand's physical motion, but \jp{コトッ} names an event that can only occur \emph{on screen}---a depiction with no off-screen referent, i.e., pseudo-haptic labeling proper. Her design critique---words sometimes switch too fast to read, are illegible during rapid motion, and the comic styling can overload the screen---directly motivates the Switch and Dens penalties of Eq.~\ref{eq:comfort} and the high-legibility display variant now shipped as a setting (Section~\ref{sec:system}).

\section{Toward Learned, Context-Aware Onomatopoeia}
\label{sec:ml}

\begin{figure}[t]
  \centering
  \includegraphics[width=\columnwidth]{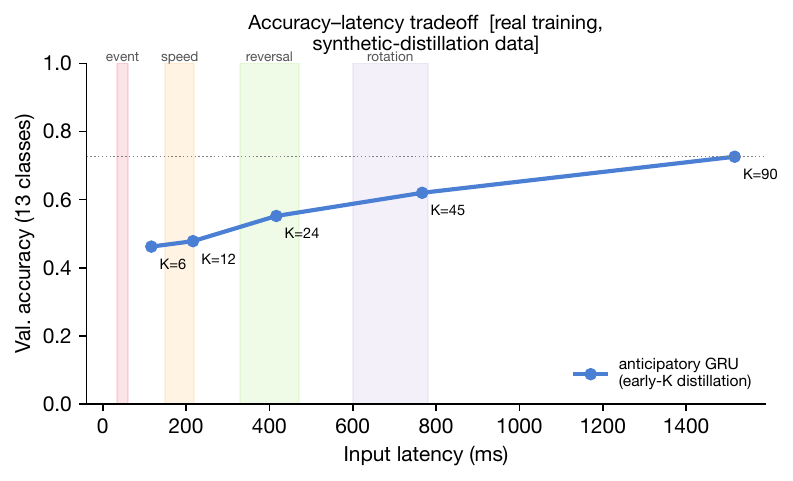}
  \caption{Accuracy--latency tradeoff of the anticipatory model: \emph{real training runs} on synthetic-distillation data (shaded bands: structural latency floors of the rule classifier). The planned context ablation is described as future work in the text and shown in no figure, as no such data exist yet.}
  \label{fig:ml}
  \Description{Line plot of validation accuracy versus input latency for the anticipatory GRU with shaded latency bands of the rule classifier.}
\end{figure}

The intended full system is a \emph{hybrid of a fast reflexive pathway and a slow contextual pathway}, arbitrated at display time---a dual-process architecture. \emph{Tier 0 (reflex, ${\approx}35$--$50$\,ms)} is the implemented event layer: raw velocity edges trigger \jp{ピタッ!!} and provisional onset words. \emph{Tier 1 (anticipation, ${\approx}120$--$220$\,ms)} is a trajectory GRU distilled from the rule classifier with training windows truncated to their first $K$ frames, predicting what a movement will be called from how it begins; it is designed to run on-device via Core~ML to refine the reflex (trained, not yet deployed). \emph{Tier 2 (context, ${\approx}0.3$--$1$\,s)} fuses the trajectory encoder with a visual embedding of a patch around the cursor, the accessibility role, application identity, and recent word history, either re-ranking the closed set or generating open-vocabulary onomatopoeia through a small language decoder (with the shared latent space of Eq.~\ref{eq:nce} supplying retrieval); all inference stays on-device so screen content never leaves the machine. The arbiter always shows the fastest available word and lets a higher tier override only when its confidence margin and semantic distance justify a switch, with hysteresis against flicker. Two components of this stack are already implemented: the OnomaFormer generator (Section~\ref{sec:onomaformer}), which stands in as an on-device open-vocabulary source, and the family-anchored constraint (Section~\ref{sec:family}) that makes such generation safe for sound-definite events. Training is staged: \emph{Phase A} distills the rule classifier (runnable today on synthetic trajectories; Figure~\ref{fig:ml}), \emph{Phase B} fine-tunes on human annotations collected in-app (the preference objective of Eq.~\ref{eq:morphloss}), \emph{Phase C} adds the contextual encoder from VLM-annotated screen recordings with human verification.

\section{Future Work: A Preregisterable Study of Agency}
\label{sec:future}

The system is built so that its own research log supplies every behavioral measure, and its display conditions are software switches. We therefore outline a within-subjects study---not yet run---to test the three formative phenomena against mundane accounts.

\textbf{Research questions.} \textbf{RQ1} (modulation): does naming an ongoing movement change explicit agency ratings and the reported felt quality of the action? \textbf{RQ2} (amplification): can animated lettering induce reported action feelings exceeding actual kinematics (e.g., \jp{ビューン} shown during slow movement)? \textbf{RQ3} (interference): does verbal mirroring feed back into motor control---changed speed distributions, increased stopping, word-seeking behavior? \textbf{RQ4} (accuracy): does correctness of the mapping matter (rule-based vs.\ random vs.\ learned)? \textbf{RQ5} (experience): effects on enjoyment, attachment, continued-use intention. \textbf{RQ6} (context): is content-aware mapping preferred over kinematics-only?

\textbf{Design.} Within-subjects conditions \{off, \textbf{scrambled} (salience-matched non-linguistic control: scrambled kana equated for luminance, size, motion energy, and onset timing), random word, mismatched word, rule-based, learned\}, counterbalanced with automated, experimenter-blind condition assignment. The \emph{word-present} contrasts are the confirmatory comparisons; off-vs-on is a manipulation check only, since participants cannot be blinded to display presence. Semantic effects are identified as differences from \emph{scrambled}, not from \emph{off}. A reading-dependence arm (non-Japanese-readers, or scrambled-kana within-subjects) tests whether effects collapse toward the salience baseline without lexical access. Display latency is additionally \emph{manipulated} (injected delays of 0/50/100/200/400\,ms) to map the temporal window of the interference effect: attention-capture accounts predict delay-insensitivity, authorship accounts predict a window. For the primary contrast (rule vs.\ scrambled on the agency composite), a power analysis targeting $d_z{=}0.65$ at $\alpha{=}.01$ (Holm-corrected) suggests $N{\approx}24$; event-level analyses use mixed-effects models. The confirmatory/exploratory split, primary DVs, and stopping rules would be preregistered (Registered Report format is under consideration). Mismatched-word trials involve mild deception and are debriefed.

\textbf{Tasks.} (a)~Free browsing; (b)~a Fitts-style pointing task (does the overlay harm pointing?); (c)~a \emph{transport task} probing the morphology channel (``place it gently'' / ``toss it,'' contrasting \jp{ポトッ} vs.\ \jp{ポーイッ！}); (d)~a \emph{click-manner task} (soft / brisk / forceful clicks, contrasting \jp{ポチ}/\jp{カチッ}/\jp{バチンッ}, and whether a mismatched form retroactively changes the felt manner of one's own click); (e)~free exploration for vocabulary discoverability. Measures: explicit SoA items, per-word-form congruence, UES-SF~\cite{obrien2018ues}, AttrakDiff~\cite{hassenzahl2003attrakdiff}, interviews, plus behavioral signatures (speed distributions, stopping probability within 300\,ms of word onset, word-transition statistics) read directly from the built-in log. A mechanism study would add event-locked kinematics, a ternary temporal-order judgment (word-first / simultaneous / stop-first) with PSS/JND, and a two-interval speed comparison for amplification on replayed probe movements---dissociating semantic meaning from induced motion cues and attentional prior entry.

\section{Ethics and Data}
\label{sec:ethics}

The released system is opt-in by construction and privacy-first. Network communication is off by default; with ``contribute data'' enabled it sends only anonymized (UUID-keyed) motion features, displayed words, and feedback to an insert-only research endpoint, after a consent dialog that names the destination and the non-recorded fields, and can be turned off at any time. Screen pixels, keystroke content, and click targets are never read; accessibility permission is used only when the corresponding mode is on, and only to obtain the UI-element role and the fact (not content) of a keypress. Each research surface has its own consent gate: research logging shows, every time it is enabled, exactly what is recorded, where, and what is excluded; and a dedicated \emph{experiment-session} mode uses a two-stage opt-in (information and consent, then participant ID), records \texttt{experiment}/\texttt{participant}/\texttt{consent} metadata, and prompts a debrief at the end---the flow the future study (Section~\ref{sec:future}) will use, including debriefing for the deceptive mismatched-word trials. Because part of the vocabulary names involuntary or affective states (\jp{プルプル} trembling, \jp{オロオロ} flustered, \jp{ドキドキ}), a system that continuously captions movement could read a disability back to its owner; we treat this as a design commitment, shipping a state-word-suppression setting, planning user-editable vocabulary, and requiring any future shared/collaborative mode to default to motion-quality words only, with state words strictly self-visible opt-in.

\section{Discussion}
\label{sec:discussion}

\textbf{Postdictive agency as a design material.} The inverted-causality report---\jp{ピタッ!!} experienced as the \emph{cause} of one's own stop, though the hand demonstrably stopped first---suggests that a system could claim authorship of user actions within the temporal binding window, modulated by semantic aptness and latency. If confirmed, this gives interaction design a new dial: not what the interface does, but \emph{who appears to have done it}. The same dial demands care; an interface that routinely absorbs authorship of user actions raises questions beyond UX.

\textbf{Morphology as a continuous interface.} Treating sound-symbolic morphology as a parameter space (Eq.~\ref{eq:morph}) turns word choice from a classification problem into a generation problem with a naturally continuous, human-annotatable output. We conjecture this is why manner-fluctuating forms feel qualitatively different from fixed event labels: the display resolves \emph{how} the action was done, which is precisely the information proprioception carries---so the annotation lands on something the body already knows, and can therefore bend it.

\textbf{Verbal haptics.} The window-edge words give a flat screen a felt topography using nothing but geometry---a haptics of vocabulary rather than actuators, extending VisualHaptics~\cite{watanabe2008visualhaptics} from pseudo-haptic cursor \emph{behavior} to pseudo-haptic \emph{language}, and complementing comic-element sensation alteration in AR~\cite{bhatia2024comics} on the unaugmented desktop. Similar verbal textures could voice scrolling friction, snapping, or system load (\jp{ヌルヌル}, \jp{もっさり}), making \emph{felt performance} inspectable. The second author's observation that \jp{コトッ} has no off-screen referent (Section~\ref{sec:vignettes}) marks these words as pseudo-haptic language in the strict sense: depictions of a contact that never physically occurred.

\textbf{Amplification, not just congruence.} The amplification log (Section~\ref{sec:amplog}) reframes the objective: the interesting words are not merely the ones that \emph{match} an act but the ones that \emph{amplify} its felt intensity. Because the signal is collected implicitly from continued action dynamics, ordinary use accrues a reward map for steering the generator---an example of an interface that improves its own expressive vocabulary from behavior alone.

\textbf{Designing from within a tradition.} We frame the Japanese grounding affirmatively rather than as a generalizability caveat: manga lettering is a centuries-refined grammar of depicting embodied motion, and gitaigo a lexicon whose morphology already parameterizes manner. Designing \emph{from within} this tradition is what yields the continuous morphology space of Section~\ref{sec:morph}; a language-neutral design would have had to invent it. The sound-symbolic \emph{dimensions} (voicing–weight, gemination–abruptness) appear cross-linguistically, so the architecture ports---each language pack re-implements $\mathcal{C}_F(\theta)$ with its own devices (consonant tensing and vowel gradation in Korean; capitalization and vowel lengthening in comic English)---while the specific forms remain proudly local.

\textbf{The OS delimits the mirror.} Our vocabulary map is also a map of macOS event visibility: cursor position is public at 60\,Hz, scrolls are globally observable, keypress facts require accessibility trust, and pinch never leaves the focused app. System-wide embodied mirroring is bounded not by sensing but by the operating system's decisions about who may see which bodily data---a constraint (and a design lever) for any future ambient-feedback system.

\textbf{From cursor to body, and to agents.} Because the pipeline verbalizes \emph{manner}, nothing binds it to the pointer: dance (a phrase's \jp{スーッ} versus its \jp{ビシッ}---whose generative inverse we have already demonstrated~\cite{okamura2023dance}), rehabilitation (supportive naming of effortful movement), and gesture are natural extensions, with the cursor study serving as the controlled miniature. The same generality applies to synthetic pointers: an auto-experiment driver already yields onomatopoeic transcripts of scripted agents, so running LLM-driven GUI agents through the system would render the \emph{texture} of machine operation legible (does an agent ever \jp{ウロウロ}?), a novel lens for human--agent comparison.

\textbf{Limitations.} The vocabulary is Japanese-first; the sound-symbolic \emph{dimensions} appear cross-linguistically, but portability of specific forms is untested. The morphology generator is a hand-tuned weak teacher pending learned replacement and human annotation, and the anticipatory tier is a feasibility check, not a demonstrated win on human data. Agency claims rest on formative self-observation ($N{=}1$, first author) and await the study of Section~\ref{sec:future}. The overlay is single-display-primary and assumes a 60\,Hz frame clock.

\section{Conclusion}

The Onomatopoeia Cursor gives the pointer a voice about the one thing it has always known and never said: the quality of our movement. It is implemented and distributed, with word \emph{forms} generated from the manner of action by a sound-symbolic morphology, an on-device transformer, and a family-anchored generator, and rendered in an evolving manga-quality lettering style. Formative use suggests the experience is immediately legible, playful, and strangely intimate---in the authors' own use, the cursor feels ``read.'' The design space, the working system, its technical evaluation, and the preregisterable study we outline lay the groundwork for understanding verbal motion mirroring as a general interaction resource.

\begin{acks}
The system is developed in the Digital Nature Group, University of Tsukuba. We thank collaborators on the onomatopoeia and sense-of-agency framing for formative discussions.
\end{acks}

\bibliographystyle{ACM-Reference-Format}
\bibliography{references}

\end{document}